\documentclass[conference]{IEEEtran}

\usepackage{cite}
\usepackage{amsmath,amssymb,amsfonts}
\usepackage{graphicx}
\usepackage{textcomp}
\usepackage{xcolor}
\usepackage{listings}
\usepackage{multirow}
\usepackage[british]{babel}
\usepackage{paralist}
\usepackage{graphics}
\usepackage{graphicx}
\usepackage{algpseudocode}
\usepackage{multirow}
\usepackage{array, booktabs}
\usepackage{xcolor}
\usepackage{xspace}
\usepackage{booktabs}
\usepackage{array}
\usepackage{ifpdf}
\usepackage{mdwmath}
\usepackage{mdwtab}
\usepackage{enumitem}

\PassOptionsToPackage{hyphens}{url}\usepackage[pdftex, colorlinks=true, hyperfootnotes=false, hyperindex=true, plainpages=false, pagebackref=false, pdfpagelabels=true, pdfstartview=FitH, linkcolor=blue, citecolor=blue, urlcolor=blue]{hyperref}

\usepackage[textsize=scriptsize,backgroundcolor=yellow!40]{todonotes}

\newcommand{\ifHideForReview}[1]{#1}
\newcommand{\ifShowForReview}[1]{}

\ifHideForReview{
  \def\toolname{Lorikeet\xspace}
  \def\laava{Laava\xspace}
}
\ifShowForReview{
  \def\toolname{MDEtool\xspace}
  \def\laava{SupplyChainStartup\xspace}
}

\begin{document}

\renewcommand{\figureautorefname}{Fig.}%
\newcommand{\subfigureautorefname}{Fig.}%
\renewcommand{\subsectionautorefname}{Section}%
\renewcommand{\sectionautorefname}{Section}%

\title{A Platform Architecture for Multi-Tenant Blockchain-Based Systems}

\author{
	\ifShowForReview{Authors hidden for review \\
     Submission 19 \\
		 ICSA 2019}
	\ifHideForReview{
	\IEEEauthorblockN{Ingo Weber, 
		Qinghua Lu, 
		An Binh Tran}
	\IEEEauthorblockA{Data61, CSIRO, Sydney, Australia\\
		firstname.lastname@data61.csiro.au}
	\and
	\IEEEauthorblockN{Amit Deshmukh, Marek Gorski, Markus Strazds}
	\IEEEauthorblockA{Laava ID Pty Ltd, Sydney, Australia\\
		firstname@laava.id}
	}
}

\maketitle

\begin{abstract}
Blockchain has attracted a broad range of interests from start-ups, enterprises and governments to build next generation applications in a decentralized manner. 
Similar to cloud platforms, a single blockchain-based system may need to serve multiple tenants simultaneously. 
However, design of multi-tenant blockchain-based systems is challenging to architects in terms of data and performance isolation, as well as scalability. 
First, tenants must not be able to read other tenants' data and tenants with potentially higher workload should not affect read/write performance of other tenants.  Second, multi-tenant blockchain-based systems usually require both scalability for each individual tenant and scalability with number of tenants.
Therefore, in this paper, we propose a scalable platform architecture for multi-tenant blockchain-based systems to ensure data integrity while maintaining data privacy and performance isolation.
In the proposed architecture, each tenant has an individual permissioned blockchain to maintain their own data and smart contracts. All tenant chains are anchored into a main chain, in a way that minimizes cost and load overheads.
The proposed architecture has been implemented in a proof-of-concept prototype with our industry partner, Laava ID Pty Ltd (Laava). 
We evaluate our proposal in a three-fold way: fulfilment of the identified requirements, qualitative comparison with design alternatives, and quantitative analysis.
The evaluation results show that the proposed architecture can achieve data integrity, performance isolation, data privacy, configuration flexibility, availability, cost efficiency and scalability.

\end{abstract}


\section{Introduction}
\label{sec:intro}

Blockchain is an emerging distributed ledger technology which has attracted a broad range of interests from start-ups, enterprises and governments \cite{aureport}\cite{ukreport} to address lack-of-trust issues in a decentralized manner. 
A large number of projects have been conducted to explore how to use blockchain to re-architect systems and to build new applications and business models.
Blockchain application areas are diverse, including supply chain, IoT, physical or digital asset registries, digital currency, payment, trade finance, and identity management. 

Similar to cloud platforms, a single blockchain-based system is often required to serve multiple tenants who reside in the same system to maintain their data. 
For example, a traceability system usually provides quality tracking services to different product manufacturers and each manufacturer can manage the tracking of their products individually.

However, design of multi-tenant blockchain-based systems is challenging to architects in terms of data and performance isolation, as well as scalability. First, tenants must not be able to read other tenants' data and tenants with potentially higher workload should not affect read/write performance of other tenants. Second, multi-tenant blockchain-based systems usually require both scalability for each tenant and scalability with number of tenants.


Therefore, in this paper, we design a scalable platform architecture for multi-tenant blockchain-based systems to achieve integrity of each tenant's data while ensuring data privacy and performance isolation. 
In the proposed architecture, each tenant has an individual permissioned blockchain to maintain their data. 
We design a custom Merkle tree in which each leaf node represents the root of each tenant's individual blockchain Merkle tree. 
We store the created custom Merkle tree on each individual blockchain and place the root of the custom Merkle tree at a pre-configured interval on a public blockchain through an anchoring component. 
This architecture design allows publicly verifiable integrity of permissioned blockchains at any time via anchoring consensus state of each permissioned blockchain at periodic intervals to a public blockchain. 
The anchoring overhead and cost remains mostly static, regardless of the number of tenants.
The architecture can also be applied to store information with different characteristics to improve flexibility and reduce cost (e.g. a long-lived blockchain and a short-lived blockchain). 

We implement the proposed architecture in a proof-of-concept prototype, as part of a collaborative project with our industry partner \laava\footnote{
\ifShowForReview{Company name hidden for peer review}
\ifHideForReview{\url{https://www.laava.id}} (accessed on 27 Nov 2018)}. 
We adopt Ethereum~\cite{buterin2013ethereum} in our implementation since it  currently offers the most mature smart contract support.
In our three-pronged evaluation, we first examine the architecture by checking the fulfilment of the identified requirements for multi-tenant blockchain-based systems. 
Second, we provide a qualitative analysis by comparing the proposed architecture with two architecture design alternatives. 
Third, we conduct a quantitative analysis by measuring the throughput under a number of conditions. 
The results show that the proposed architecture can achieve data integrity, performance isolation, data privacy, configuration flexibility, availability, cost efficiency and scalability.

\begin{figure}[t]
	\begin{center}
		\center
		\includegraphics[width = \columnwidth]{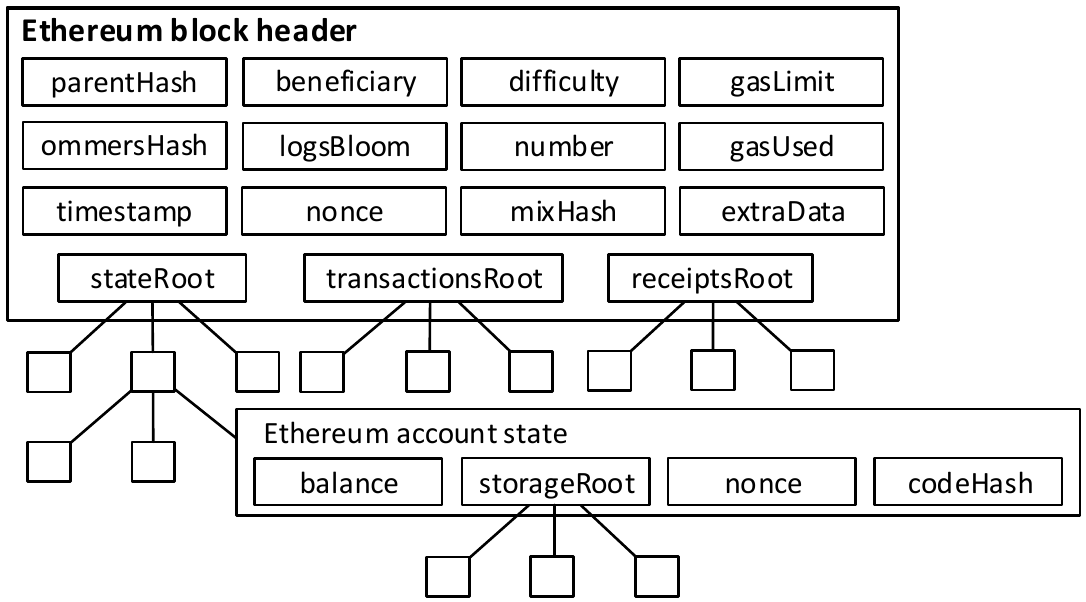}
		\caption{Ethereum block header and state merkle tree.}
		\label{tree}
	\end{center}
\end{figure}

The remainder of this paper is organized as follows. 
The next section discusses background and related work. 
\autoref{sec:reqs} introduces the requirements for multi-tenant blockchain-based systems. 
\autoref{sec:arch} presents the proposed architecture. 
\autoref{sec:use-case} discusses the prototypical implementation of the architecture in the context of \laava's use case. 
\autoref{sec:eval} evaluates the proposed architecture before
\autoref{sec:concl} concludes the paper and outlines the future work.

\section{Background and Related Work}

\subsection{Background: Blockchain}

A \emph{blockchain} is a distributed append-only store of transactions distributed across computational nodes and structured as a linked list of blocks, each containing a set of transactions%
\ifHideForReview{~\cite{2019-Blockchain-Book}}%
\ifShowForReview{~\cite{anonymous}}.
Blockchain was introduced as the technology behind Bitcoin \cite{Satoshi:bitcoin}. 
Its concepts have been generalized to \emph{distributed ledger} systems that verify and store any transactions without coins or tokens \cite{scheuermann2015iacr}, without relying on any central trusted authority like traditional banking or payment systems. 
Instead, all participants in the network can reach agreements on the states of transactional data to achieve trust.


\emph{Merkle trees} are an important part of blockchain, supporting fundamental blockchain functionality and enabling efficient and secure verification of large data structures. 
Merkle trees have a hash-based structure that can ensure data integrity in a trivial way: each node (except leaves) in the tree contains the hash of its child node values;
if nothing changed, the root will be the same; otherwise only the hashes on the path from the root to the changed leaves are changed.
The Merkle tree used in the Ethereum blockchain platform is called Merkle Patricia tree \cite{buterin2013ethereum}. 
There are three different Merkle Patricia tree structures in Ethereum, as illustrated in \autoref{tree}: state tree, transaction tree and receipt tree. 
Every block header contains the roots of those three trees.
The global state tree contains a key-value pair for every account in the Ethereum network and is updated by every transaction.
The key is the account address while the value is an encoding of details including nonce, balance, storageRoot and codeHash.
The root of state tree is cryptographically dependent on all state tree data and can be used as a unique and secure identifier for the state tree.

A smart contract is a user-defined program that is deployed and executed on a blockchain system \cite{Omohundro:2014\ifHideForReview{,2019-Blockchain-Book}}, which can express triggers, conditions and business logic \cite{Weber:BPM2016} to enable complex programmable transactions. 
Smart contracts can be deployed and invoked through transactions, and are executed across the blockchain network by all connected nodes. 
The signature of the transaction sender authorizes the data payload of a transaction to create or execute a smart contract. 
Trust in the correct execution of smart contracts extends directly from regular transactions, since (i) they are deployed as data in a transaction and thus immutable; (ii) all their inputs are through transactions and the current state; (iii) their code is deterministic; and (iv) the results of transactions are captured in the state and receipt trees, which are part of the consensus. 

When using a blockchain, there are different types of deployments, including public blockchain, consortium blockchain or private blockchain. 
Public blockchains, which can be accessed by anyone on the Internet (\emph{``permission-less''}), have high information transparency and auditability, but sacrifice performance and a cost/incentive model. 
A consortium blockchain is typically used across multiple organisations and the rights to read/write on the blockchain may be restricted to specific participants.
In a private blockchain network, write permissions are often kept within one organisation, although this may include multiple divisions of a single organisation.
Private blockchains are the most flexible for configuration because the network is governed and hosted by a single organisation.
A blockchain may be \emph{permissioned} in requiring that one or more authorities act as a gate for participation. 
This may include permission to join the network and read information from the blockchain, to initiate transactions, or to create blocks. 
Permissions can be stored either on-chain or off-chain.
There are often tradeoffs between permissioned and permission-less blockchains including transaction processing rate, cost, censorship-resistance, reversibility, finality and flexibility in changing and optimising the network rules. 

\subsection{Related Work}
There are a number of projects which have been conducted to address blockchain limitations including scalability, privacy and cost. 
Quorum\footnote{\url{https://www.jpmorgan.com/global/Quorum}} addresses specific challenges to blockchain technology adoption in the financial industry, which supports both public and private smart contracts to enable data privacy.
Plasma \cite{poon2017plasma} is designed to be scalable to a large amount of state updates by providing incentivised and enforced execution of smart contracts via transaction fees.
The Dfinity blockchain \cite{DFINITYWhitePaper} provides a scalable consensus mechanism which can scale through continuous quorum selections driven by a random beacon. In Dfinity, the interblock time (interval between two blocks) takes a few seconds and a transaction is committed after only two confirmation blocks.
Komodo\footnote{\url{https://komodoplatform.com/}} includes a delayed Proof of Work consensus mechanism to ensure security while avoiding direct competition. 
Stellar\footnote{\url{https://www.stellar.org/}} provides a distributed payment infrastructure, which takes 2-5 seconds to reach consensus. 
EOS\footnote{\url{https://eos.io/}} is designed to enable vertical and horizontal scaling of decentralized applications by providing an operating system-like construct, which can handle to thousands of transactions per second without fees.

Many efforts have considered the area of multiple blockchains and sides chains. 
Kan et al. \cite{LuoKan2018} propose an architecture for reliably exchanging information across multiple blockchains. A connection model is designed for routing management in multiple blockchains, which can provide atomicity and consistency for transactions across blockchains and allows increasing throughput. 
Cash and Bassiouni \cite{Cash2018} propose a two-tier blockchain architecture that utilizes a permission-less tier for decentralization and security, and a centralized tier that focuses on data control and restrictions. In the architecture, tier one allows any node to read from and write to the blockchain, while tier two only allows restricted users to process read and write operations. 
The Loom Network\footnote{\url{https://loomx.io/}} is a scaling solution for Ethereum, which provides a network of Delegated Proof of Stake (DPoS) sidechains allowing for highly-scalable decentralized applications while still being backed by the security of Ethereum. 

Supply chain and registries are two promising areas for applications of blockchain. 
Most of the existing work on supply chain \cite{Qinghua2017,AgriSupplychain,KSHETRI201880,Kim2018,Wu2017A} focuses on designing blockchain-based systems to achieve item traceability by leveraging the fundamental properties of blockchain. 
Lu and Xu~\cite{Qinghua2017} shared the experience of building originChain, an adaptable blockchain-based system which provides transparent tamper-proof traceability data and automates regulatory-compliance checking. 
Tian \cite{AgriSupplychain} combines Radio-Frequency Identification (RFID) and blockchain technology to  build a food supply chain traceability system, which covers the whole traceability management process for quality and safety of food. 
Kim and Laskowski \cite{Kim2018} analyse a traceability ontology and translate some of its representations to smart contracts that execute a provenance trace and enforce traceability constraints. 

Building registries on a blockchain can guarantee data integrity, availability, transparency and immutability, which are key requirements for registries \cite{Downey2016}.
There are registries being built on blockchain in ad-hoc ways, for example, Namecoin\footnote{\url{https://namecoin.org/}}, which is a domain name registry that shares the same network with Bitcoin\footnote{\url{https://bitcoin.org/}}, and Abscribe\footnote{\url{https://www.ascribe.io/}}, which is an artwork registry that allows artists to register and manage the ownership of their digital artwork. 
However, building a registry on blockchain is non-trivial due to the steep learning curve of the technology. 
Regis\footnote{\url{https://regis.nu/}} is a contract generator on Ethereum\footnote{\url{https://www.ethereum.org/}} blockchain, but only provides very basic operations. 
A registry generator for blockchain was introduced in a demo paper~\cite{Tran2017}.

However, we are not aware of any work addressing the challenges of commercial multi-tenant systems on blockchain, such as performance isolation and data privacy.

\begin{figure*}[t]
	\begin{center}
		\center
		\includegraphics[width = .9\textwidth]{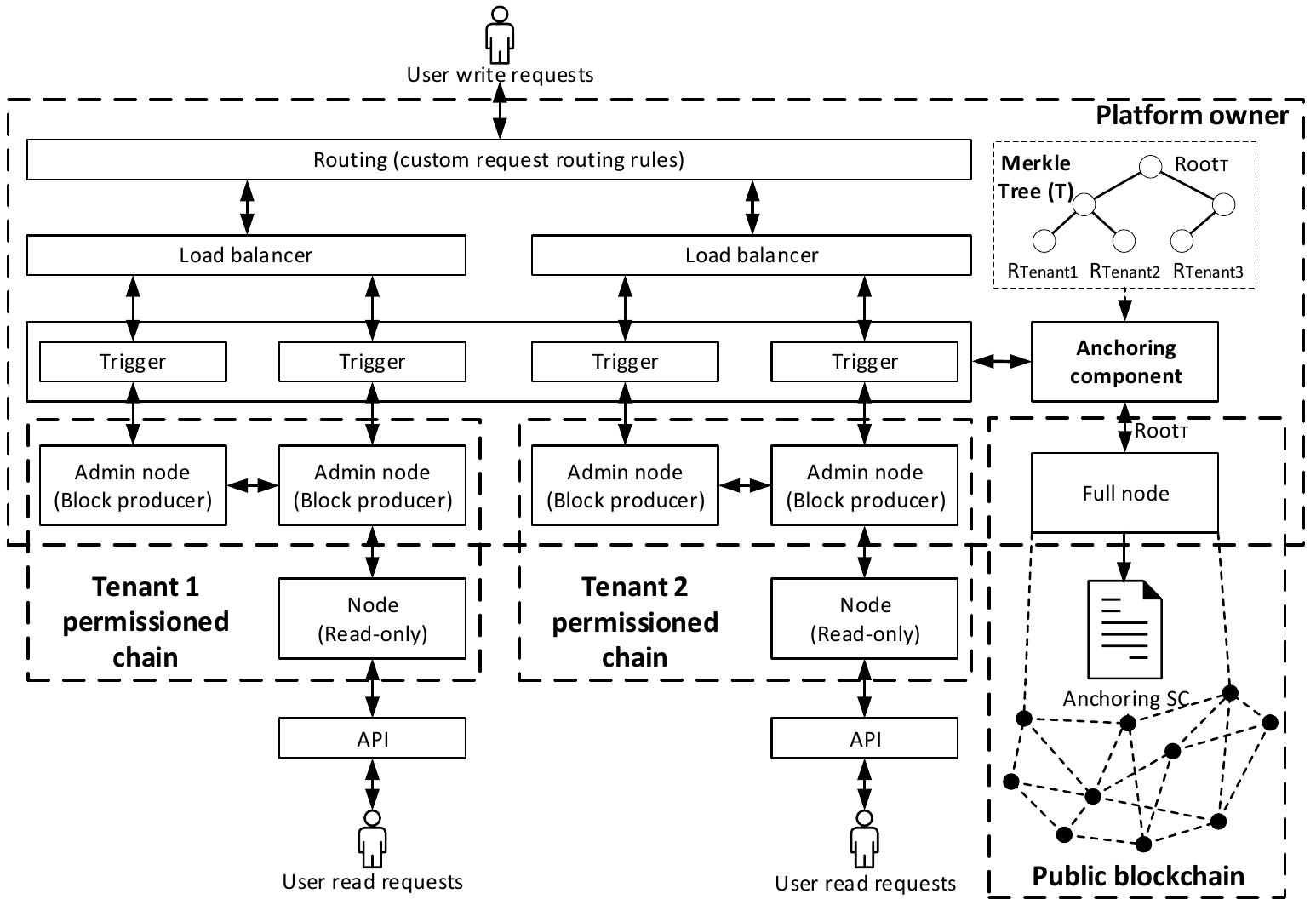}
		\caption{Platform Architecture for multi-tenant blockchain-based systems.}
		\label{architecture}
	\end{center}
\end{figure*}

\section{Requirements}
\label{sec:reqs}

We gathered application-agnostic functional and non-functional requirements of multi-tenant blockchain-based systems, which are described below.
%
We followed standard requirements elicitation methodologies \cite{Kotonya:1998:REP:552009} in our work with \laava for their specific requirements. Subsequently we abstracted and filtered these system-specific requirements to derive a list for the more general class of multi-tenant systems. Some of the requirements here might differ from the needs of other systems, while others might need to be refined. Therefore, the list can be viewed as assumptions and drivers for the architecture we discuss in the rest of the paper. If applied to another system, the changes to the requirements may need to be reflected in an adaptation of the architecture.

\subsection{Functional Requirements}






\noindent\textit{FR1 -- Writing data on blockchain restricted to selected clients:}
The platform shall have the ability to write data on blockchain, restricted, e.g., to the platform owner.

\noindent\textit{FR2 -- Writing batches of data on blockchain:}
The platform shall have the ability to write batches of data with low cost and overhead.

\noindent\textit{FR3 -- Viewing the entire history:}
The platform shall have ability to read the entire data history, i.e., all historical events and data values over time.



\noindent\textit{FR4 -- Tracking authenticity of data:}
end users need to be able to see and validate the identity of clients that wrote data to the system.

\noindent\textit{FR5 -- Providing external auditing/verification for independent agencies:}
independent agencies need to be able to access data for auditing for each individual tenant.

\noindent\textit{FR6 -- Providing a multi-tenant platform:}
the platform supports multiple tenants to serve their end users, where different tenants can have different business needs.

\subsection{Non-Functional Requirements}
\noindent\textit{NFR1 -- Data integrity}
of on-chain data must be ensured.

\noindent\textit{NFR2 -- Scalability:} 
\begin{itemize}[noitemsep,topsep=0pt]
	\item Scalability within each tenant. For example, a tenant might store large amounts of data within a period of time.
	\item Scalability in the number of tenants.
\end{itemize}

\noindent\textit{NFR3 -- Data Privacy:} 
in general, tenants must not be able to read other tenants' data (e.g., how many unique item IDs were created, scan event counts, timing, or locations).

\noindent\textit{NFR4 -- Performance Isolation:} 
tenants with potentially higher workload (e.g., commodity goods with millions of events daily) should not affect read/write performance for other tenants.
	
\noindent\textit{NFR5 -- Availability:} 
the blockchain infrastructure must be available, in terms of responsiveness to read/write operations.

\section{Architecture}
\label{sec:arch}

In this section, we propose a platform architecture which can meet the above requirements of multi-tenant blockchain-based systems.
\subsection{Overall Architecture}

\begin{figure}[t]
	\begin{center}
		\center
		\includegraphics[width = 0.4\textwidth]{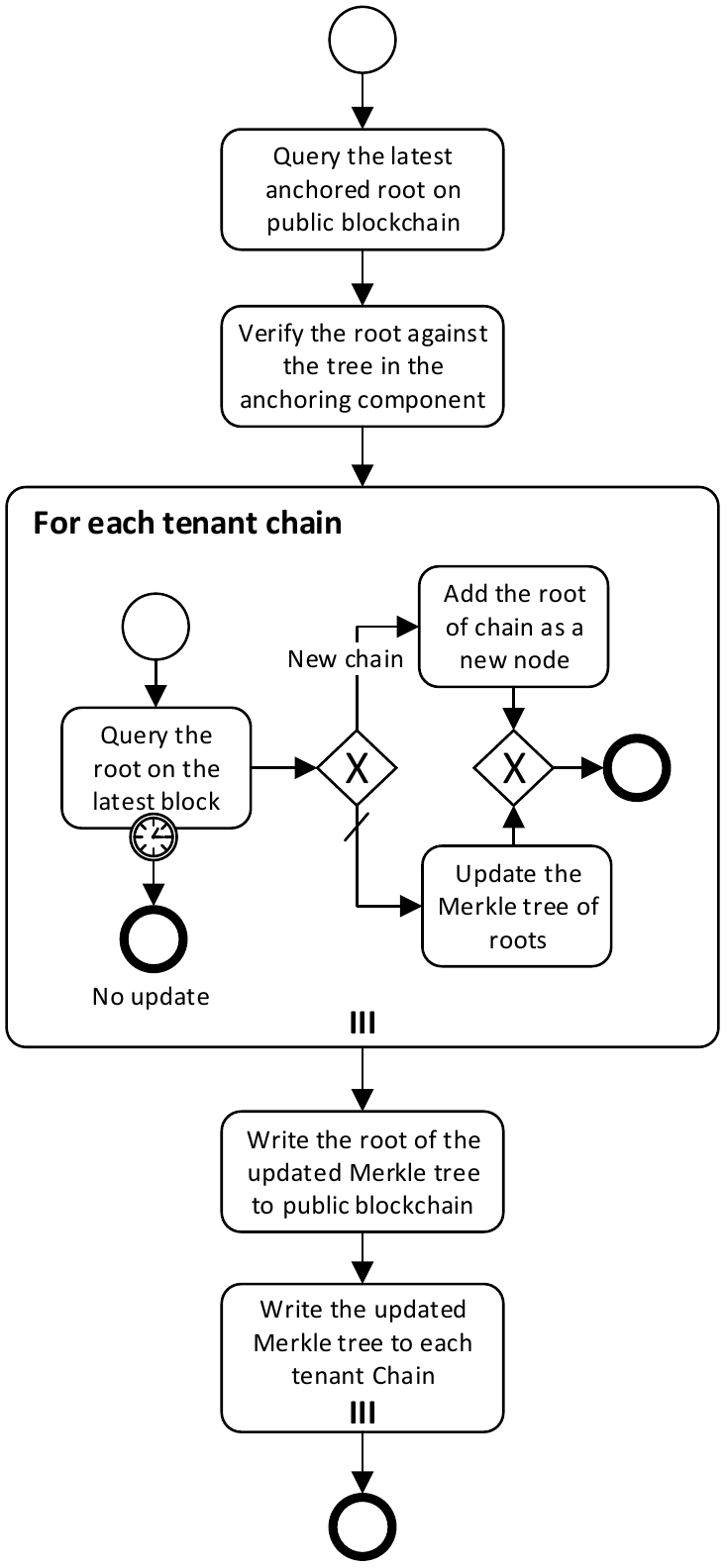}
		\caption{The anchoring protocol. (Notation: BPMN)}
		\label{protocol}
	\end{center}
\end{figure}

\autoref{architecture} illustrates the platform architecture we propose for multi-tenant blockchain-based systems.
Each tenant has an individual permissioned blockchain to maintain their information. 
The platform owner hosts all admin nodes for producing blocks while the tenants/auditors host read-only nodes, which can be enforced through the blockchain configuration. 
The platform owner has access to all the tenants' on-chain data, and provides APIs for writing to tenants' chains.

An end user, e.g., a client of a tenant, sends write requests through the routing component, which forwards them to respective tenant's blockchain to process. 
The load balancer distributes the workload to the trigger components for different admin nodes of the same tenant. 
The functionality provided by the trigger includes writing data to the blockchain and communicating with the anchoring component.
The user sends read requests to the read-only node through a public API. 

The proposed architecture allows publicly verifiable integrity of private blockchains at particular times via anchoring the consensus state of each private chain at periodic intervals to the public blockchain.
The anchoring component connects to a node in the public blockchain network, and one node for each tenant's blockchain to be anchored.
We design a Merkle tree $T$, as shown in the top right corner of \autoref{architecture}, in which each leaf node represents the root of each tenant blockchain Merkle tree $R_{TenantX}$. 
We store the newly created Merkle tree $T$ on each individual blockchain and place its root $Root_{T}$ on a public blockchain through the anchoring component. 
The architecture can also be applied to store information with different characteristics to improve flexibility and reduce cost (e.g. a long-lived blockchain and a short lived blockchain). 

There are two smart contracts in this architecture: a smart contract in each tenant's permissioned blockchain and a smart contract in the public blockchain.  The smart contract in each tenant's blockchain is pre-deployed and included as part of genesis block. The Merkle tree data structure for $T$ is stored in this smart contact, while its root $Root_{T}$ is placed in the smart contract in the public blockchain.
The Merkle tree implementation uses Ethereum's Merkle-Patricia tree library.

This design assumes that the platform owner can be relied on by tenants in handling the anchoring process, which leads to the design decision that only the platform owner is hosting the anchoring component. 
Tenants can continuously monitor that the platform owner is performing the anchoring process in a correct manner, as described below. 
In other words, the trust in the platform owner only extends to it \emph{performing} the anchoring, no trust in its \emph{correctness} is required.

\subsection{Anchoring protocol}
The anchoring scheduler is configured as agreed between the platform owner and tenants (e.g. every 10 minutes). 
The identity of each blockchain is established using the hash of genesis block. 
\autoref{protocol} describes the way the anchoring protocol works at anchoring time, in the BPMN notation~\cite{omg2013bpmn}. 

The protocol starts with querying the latest anchored Merkle root stored on the public blockchain and verifying the Merkle root against the tree maintained in the anchoring component to make sure the anchoring component is up to date. 
For each tenant's permissioned blockchain registered with the anchoring component, 
there is a subprocess in \autoref{protocol}; all subprocesses are executed in parallel (see marker ``III'' at the bottom).
In such a subprocess, say for tenant $X$, the protocol queries the blockchain Merkle roots on the latest block. If the chain is in the Merkle tree of roots, the value for $R_{TenantX}$ is updated. 
If not, e.g. $X$ is a new tenant, the $R_{TenantX}$ is added as a new node of the Merkle tree of roots. 
The tenant blockchain node might not be available, 
and the request times out. In that case, this tenant chain's root is not updated.

After all tenant blockchain roots are processed, the protocol writes the Merkle root $Root_T$ of the updated Merkle tree of roots to public blockchain along with the previous root, and stores the content of the updated tree to the smart contract pre-deployed on each tenant's blockchain.

The transaction that anchors the Merkle root to public blockchain might take time to be included and committed, which may be longer than the anchor interval. 
Thus, the anchoring scheduler is using a simple lock-based mechanism. Whenever a new anchoring round starts, it claims the lock. 
New anchoring rounds are always scheduled according to the interval. 
When the next round is scheduled, it will first check whether the lock is available. 
If the lock is not available, then that round will be skipped.

\subsection{Auditing process}

The integrity auditing process is as follows.
\begin{itemize}[noitemsep,topsep=0pt]
	\item The auditor needs to run a node of a tenant's blockchain (say, Tenant $X$) as the auditor node.
	\item The auditor needs to read the latest anchoring point on public blockchain and obtain the Merkle root $Root_{T}^{'}$ and the Merkle root of the corresponding block of this tenant's blockchain (i.e. $R_{TenantX}^{'}$).
	\item The auditor compares $R_{TenantX}^{'}$ with the root stored in the Merkle tree of roots for Tenant $X$'s blockchain at anchoring time (i.e. $R_{TenantX}$).
	\item The auditor compares $Root_{T}^{'}$ with the value of the Merkle tree stored in the tenant's blockchain (i.e. $Root_{T}$).
\end{itemize} 

By performing the auditing process, the auditor can continuously monitor the data written to each tenant chain and the correctness of anchoring performed by the platform owner.

\section{Use Case and Implementation}
\label{sec:use-case}

\subsection{Use Case}

Product counterfeiting and fraud are costly for the industry and potentially dangerous for customers, especially if medicine and food products are affected. 
These problems are wide-spread across many industries and supply chain processes. 
\laava is a third-party item tracking service provider which provides a novel type of unique ID for individual item tracking with various interesting features. The unique IDs are designed in a way that makes counterfeiting harder, and has numerous advantages over barcode and QR codes. 
Regarding system requirements, hundreds of product manufacturers may become tenants and use the system to manage their products' traceability information, and millions of product consumers may use it to access the information. 
\laava clients will create unique IDs on a blockchain at the point of packaging or production. 
The products with individual unique IDs on them will flow through the supply chain and pass through multiple points of scanning until they reach consumers. 

\ifHideForReview{The authors from Data61, CSIRO, }
\ifShowForReview{The authors from a research organization} developed a prototype  for the \laava use case in a collaborative project with product managers, architects, and developers from \laava.
In particular, the project seeks to allow any individual physical or digital thing to be authenticated easily and securely, using blockchain.

\subsection{Implementation}
We used a model-driven engineering tool called \toolname \cite{Tran2018LorikeetAM}, which can automatically produce smart contracts from business process models and registry data schema, to implement the proposed architecture design. The smart contracts are written in Solidity, compiled with Solidity compiler version 0.4.24. We used Truffle framework\footnote{\url{https://truffleframework.com}} to compile and test smart contracts. The trigger and anchoring components are written in TypeScript with Node.js version 10, implementing the REST API using express.js server.
Blockchain miners order pending transactions first by account nonce, then gas price for inclusion to new blocks. Thus, in each tenant's permissioned blockchain trigger, we used a different Ethereum account and provided higher gas price for anchor-specific transactions. This is to make sure the anchor transactions are not delayed (having separate nonce) and have highest priority to be included in each tenant's chain. 
\autoref{data} shows the on-chain data structure designed for anchoring in multi-tenant blockchain-based systems, which includes BlockchainIDOracleInterface, SerializedTrieStore, and PublicAnchor. 

\begin{figure}[t]
	\begin{center}
		\center
		\includegraphics[width=0.48\textwidth]{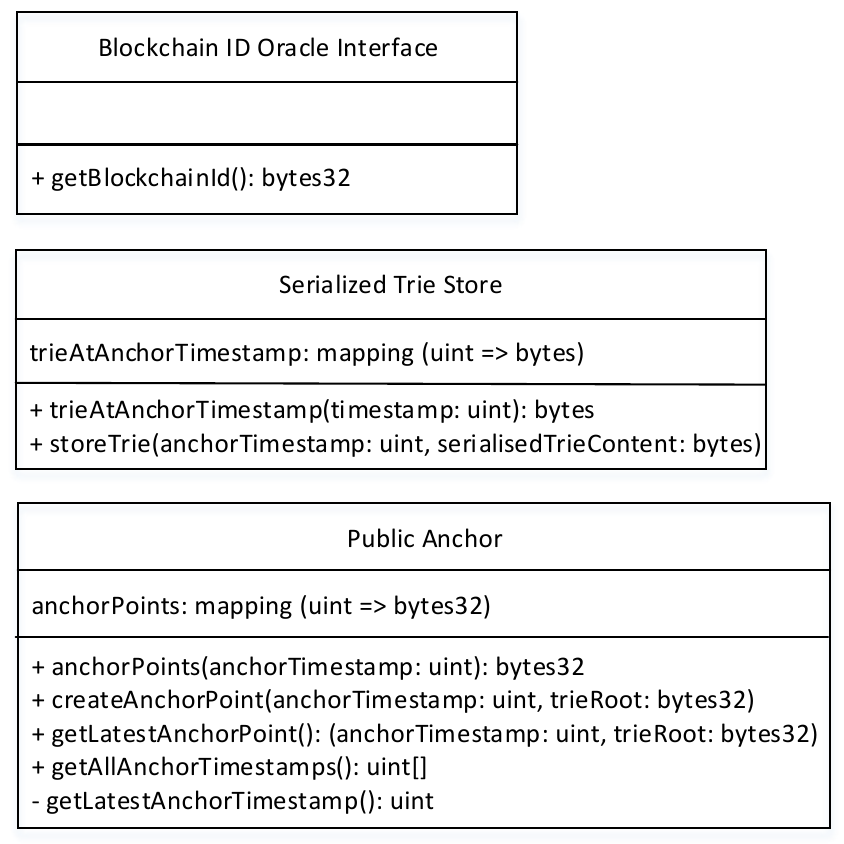}
		\caption{On-chain data structure for anchoring}
		\label{data}
	\end{center}
\end{figure}

\section{Evaluation}
\label{sec:eval}

In this section, we evaluate the proposed architecture design in terms of requirements fulfilment, qualitative analysis and quantitative analysis.
%
For requirements fulfilment, we examined the implemented proof-of-concept using the proposed architecture against the functional and non-functional requirements identified in \autoref{sec:reqs}.

\subsection{Functional Requirements Fulfilment}
\textit{FR1 - Writing data on blockchain restricted to selected clients}
The tenants are able to register unique ID, meta data, and scan events on-chain via the API provided by \laava.

\textit{FR2 - Write batches of data on blockchain}
We implemented a function in the unique ID registry contract for creating an array of unique IDs, so that a batch of multiple unique IDs can be registered via one function call, i.e. one blockchain transaction.

\textit{FR3 - View the entire history}
The tenants are able to read all historical events and data values over time via the API provided by \laava.

\textit{FR4 - Track authenticity of data}
Once a consumer scans a unique ID, they are able to validate the identity of the manufacturer who registered the unique ID.

\textit{FR5 - Independent agencies to provide external auditing/verification}
The independent agencies are able to access data for auditing for each individual tenant via the read-only nodes hosted by the independent agencies.

\textit{FR6 -- Multi-tenant platform}
The  platform supports multiple tenants to have individual permissioned blockchain to serve their end users, where different tenants can have different business needs.

\subsection{Non-Functional Requirements Fulfilment}
\textit{NFR1 - Data integrity}
Data integrity is achieved via anchoring to public blockchain.

\textit{NFR2 - Scalability} 
The operations of registering unique ID and scan event etc. are on permissioned blockchain so there is no transactional cost involved (compared to public blockchain). The cost mainly involves maintaining the infrastructure for permissioned blockchains to ensure availability, which shows good scalability within one tenant.

The cost for anchoring to public blockchain is fixed since only the combined Merkle root of all tenant chains' Merkle roots are written to public blockchain at predetermined intervals. Thus, the proposed design is scalable in respect to the number of tenant chains.

\textit{NFR3 - Data Privacy} 
Data privacy is enabled since each tenant has an individual tenant chain which has restrict permissions to join and runs on separate networks (i.e. VPCs).

\textit{NFR4 - Performance Isolation} 
Tenants have own permissioned public blockchains. Thus, transactions on one chain would not affect others. 

\textit{NFR5 - Availability} 
Each tenant chain maintains sufficient replication to ensure availability for each tenant's chain. 

\subsection{Qualitative Analysis}
In this section, we evaluate three design alternatives against the identified non-functional requirements listed in \autoref{sec:reqs}. 
\subsubsection{Design Alternatives}
We evaluate the architecture design by comparing three design alternatives: architecture using public blockchain (Design Alternative 1 illustrated in \autoref{alternative1}), architecture using global chain anchoring to public blockchain (Design Alternative 2 shown in \autoref{alternative2}), architecture using multiple blockchains anchoring to public blockchain (Design Alternative 3 -- the proposed architecture design discussed in \autoref{sec:arch}).

As shown in \autoref{alternative1}, in Design Alternative 1, all the information is stored on public blockchain. The public blockchain provides a neutral data store to maintain unique ID information. Anyone on the Internet can access the unique ID information stored on the public blockchain using the deployed smart contracts. The platform owner's existing backend communicates with the unique ID registry smart contracts deployed on blockchain via the blockchain trigger. The unique ID registry smart contracts are deployed on the public blockchain network. The blockchain trigger and local blockchain node are hosted on one virtual machine (VM). The platform owner's existing backend interacts with the blockchain trigger via REST API. In the blockchain trigger, there are two layers: blockchain communication layer and business logic layer. Blockchain communication layer consists of three components, including sending blockchain transactions, querying smart contract states, and listening to transaction progress and smart contract events. Sending blockchain transactions processes write operations while querying smart contract states focuses on read operations. Blockchain trigger obtains status of transactions and receives smart contract events via the listening to transaction progress and smart contract events component. The business logic layer comprises different business logic for each REST API, which is processed through the blockchain communication layer. 

The Design Alternative 2 is illustrated in \autoref{alternative2}. Similar to Design Alternative 3, the anchoring schedule is time-based (e.g. every 10 mins), which is configured and agreed between the platform owner and tenants. The anchoring component stores the Merkle root of global blockchain ($Root_{GlobalChain}$) to the public blockchain we are anchoring to, together with the block number and block hash of global blockchain at anchoring time. To audit the data integrity of global blockchain, the auditor runs a node of the global blockchain and read latest anchor point on public blockchain. Then the auditor compares $Root_{GlobalChain}$ at anchoring time with the information stored on public blockchain.

\begin{figure*}[t]
	\begin{center}
		\center
		\includegraphics[width = 0.9\textwidth]{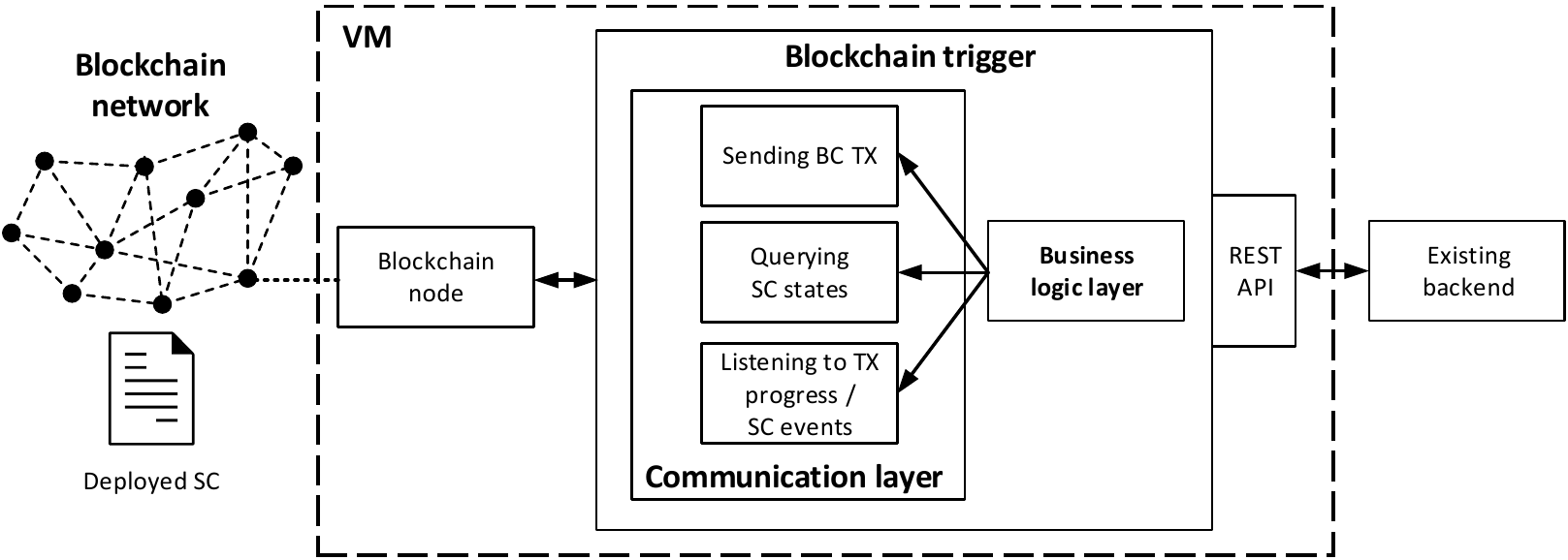}
		\caption{Design alternative 1 of multi-tenant blockchain-based systems.}
		\label{alternative1}
	\end{center}
\end{figure*}

\begin{figure*}[t]
	\begin{center}
		\center
		\includegraphics[width = 0.7\textwidth]{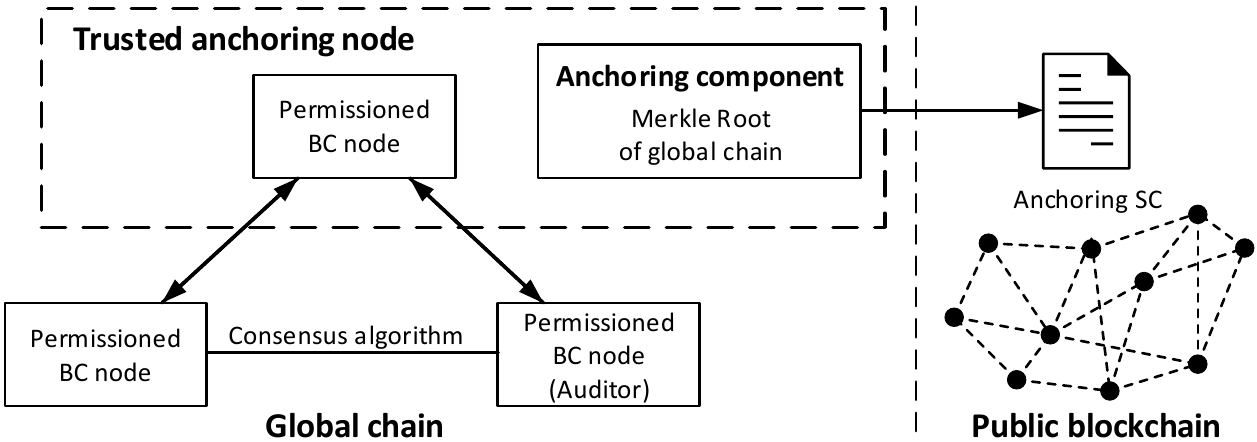}
		\caption{Design alternative 2 of multi-tenant blockchain-based systems.}
		\label{alternative2}
	\end{center}
\end{figure*}

\subsubsection{Data Integrity}
Data integrity is achievable by using all the three design alternatives. 
In all the three design alternatives, creating a unique ID registry entry is done by \laava in the current implementation. 
In Design Alternative 1, all the tenants as blockchain network participants hold a local copy of the blockchain, through which they can access the unique ID registry on blockchain.
In Design Alternative 2 and Design Alternative 3, data integrity is guaranteed via anchoring to public blockchain. Design Alternative 2 stores the Merkle root of the global blockchain to the public blockchain while Design Alternative 3 keeps the root of the Merkle roots of each tenant's blockchain on the public blockchain. 

\subsubsection{Cost}
Both Design Alternative 2 and Design Alternative 3 are designed in a way that anchors to public blockchain. The cost for anchoring to public blockchain is fixed as only Merkle root of global blockchain (Design Alternative 2) or the combined Merkle root of all tenant blockchain's Merkle roots (Design Alternative 3) are written to public blockchain at pre-determined interval. Regarding infrastructure cost, platform owners only needs to host one node for global blockchain in Design Alternative 2, while the platform owner must host at least one node for each individual tenant's blockchain in Design Alternative 3. To maintain availability, potentially higher cost is needed with the increased number of tenants. 

\subsubsection{Data Privacy}
Tenants are required to read their own product data but not for competitors data. Data is encrypted before storing to public blockchain in Design Alternative 1 and to global blockchain in Design Alternative 2. Design Alternative 3 restricts the ability to join individual tenants' blockchain as each has different genesis block and chain ID. Also, in Design Alternative 3, nodes from each tenant's blockchain run on separate virtual private clouds (VPCs).

\subsubsection{Performance Isolation}
In Design Alternative 1 and Design Alternative 2, tenants with higher transactional volume and throughput might affect performance for lower-throughput tenants since all the data are written through one blockchain trigger. In Design Alternative 3, transactions on one chain would not affect others since tenants have own permissioned blockchains and each blockchain has its own trigger for writing data to the corresponding blockchain.

\subsubsection{Availability}
Design Alternative 1 can achieve availability since it uses public blockchain. 
Both Design Alternative 2 and Design Alternative 3 can increase availability by adding more full nodes and block producers. 
Infrastructure cost and maintenance overhead may increase with number of tenants. 
Design Alternative 1 needs overall less number of replication nodes as all tenants use one global blockchain. 

\subsubsection{Configuration Flexibility}
Both Design Alternative 2 and Design Alternative 3 can be independent of particular blockchain forms. Different blockchains with different consensus algorithms can be used for permissioned blockchain. Also, both design can anchor to different public blockchains which do not necessarily need to support smart contracts.

In Design Alternative 2, all tenants need to agree on using the same blockchain platform, consensus algorithm and configuration while Design Alternative 3 has flexibility to choose different blockchain platforms (e.g. can use Hyperledger Fabric for a particular tenant and use Ethereum for others), consensus algorithms and blockchain configurations (e.g. inter-block time) for each tenant. Only anchoring protocol need to be agreed by all tenants.

\subsection{Quantitative Analysis -- Performance and Scalability}

\begin{figure*}[t]
	\begin{center}
		\center
		\includegraphics[width = 0.7\textwidth]{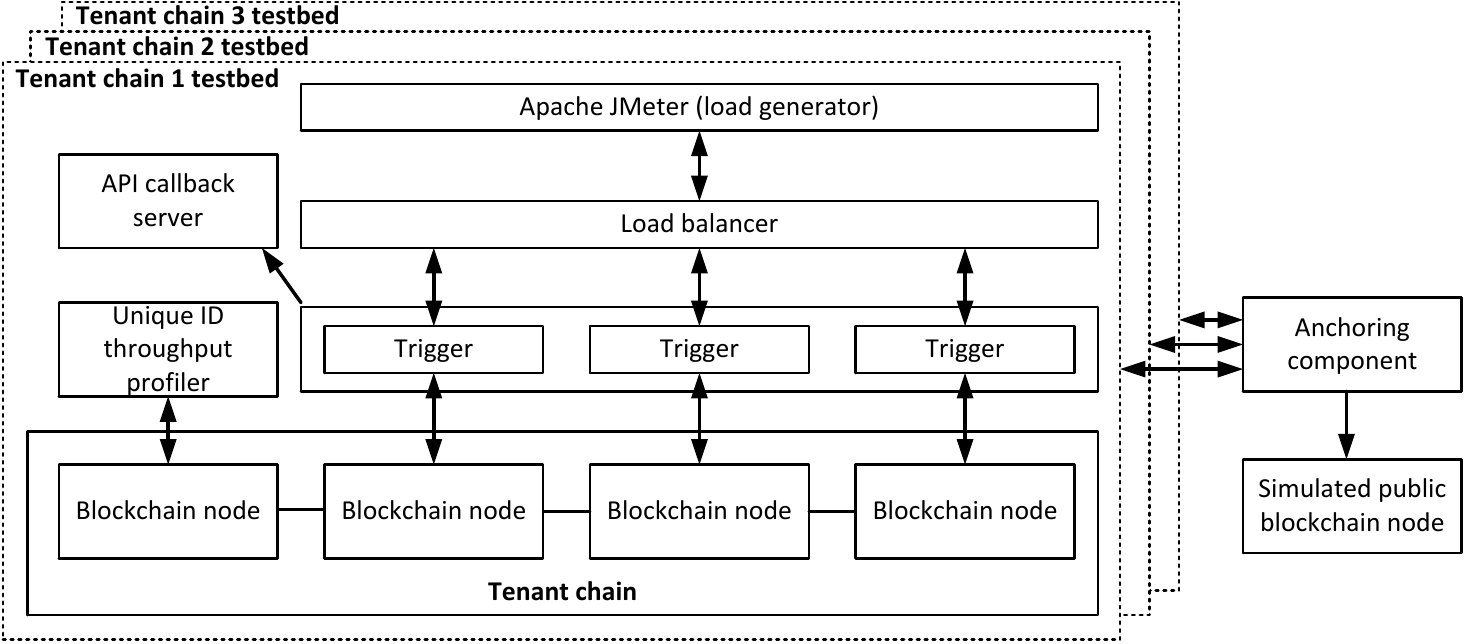}
		\caption{Deployment architecture for quantitative analysis experiments.}
		\label{experiment}
	\end{center}
\end{figure*}

There are two objectives for the quantitative analysis. The first objective is to measure the unique ID creation throughput since the unique ID creation is one of the most important functional requirements of the use case. The second objective is to evaluate the anchoring process, since anchoring performance is critical for the feasibility of the overall architecture design. In particular, the anchoring protocol needs to operate regardless of the application load on the tenant chains.

\subsubsection{Experiment design}
\autoref{experiment} shows the experiment deployment architecture for measuring the performance and scalability of the prototype implementing the proposed architecture. 

Components in the experiment design were deployed as Docker containers on AWS\footnote{\url{https://aws.amazon.com/}} EC2 virtual machines. 
We deployed the anchoring component on a dedicated m5.xlarge EC2 instance (4 vCPUs, 16 GB RAM, 20GB EBS disk), which communicates with all tenant chain testbeds.
Each tenant chain testbed used 
(i) 4 m5.xlarge EC2 instances for blockchain nodes, triggers, and other components; 
(ii) an Application Load Balancer (AWS ALB); and
(iii) one m5.2xlarge instance (8 vCPUs and 32 GB RAM) for the JMeter load generator.


The permissioned tenant chain uses the Ethereum client \emph{Parity} and its Proof-of-Authority (PoA) implementation\footnote{\url{https://wiki.parity.io/Proof-of-Authority-Chains}}, and has 3 authorities (i.e. block producing nodes) which are connected to a trigger each. There is also one read-only node connected to both the permissioned chain and the transaction profiler. 
The block-producing nodes use different authority accounts. 
Blocks are only produced when there are pending transactions. 
The anchoring component is connected to a simulated public blockchain node with an inter-block time of 15 seconds, which is approximately the median for public Ethereum\footnote{\url{https://blog.ethereum.org/2014/07/11/toward-a-12-second-block-time/}}.

The load generation throughput is produced via JMeter, which requests the creation of a high amount of new unique IDs by calling the respective API. 
It is configured to 20 creations per batch (API call \& blockchain transaction). The test duration is 1 hour.
The block gas limit in the tenant chain is set to 80M gas and the inter-block time is configured to 5 second (the minimum recommended for Parity PoA\footnote{\url{https://github.com/paritytech/parity-ethereum/issues/9586}}). 
Each batch transaction consumes 1.05 million gas. Therefore, at most 76 transactions fit into a block, limiting the theoretical maximum throughput to 15.2  transactions per second (tps) -- corresponding to 304 unique ID creations per second.
%
We run four different tests to measure the transaction sending and inclusion throughput over time with different loads:

\textbf{Test 1}: normal load scenario ($<$15tps), one tenant chain.

\textbf{Test 2}: boundary load scenario (starting at $\approx$18tps), one tenant chain. 

\textbf{Test 3}: overload scenario ($\approx$18-25tps), one tenant chain, i.e., the incoming throughput is higher than the theoretical throughput limit of the blockchain. 

\textbf{Test 4}: an overload scenario with three tenant chains ($\approx$18-25tps on each chain), i.e., Test 3 on three tenant chains in parallel.
With this test we investigate the performance isolation between tenant chains as well as the anchoring protocol. 

The actual blockchain transaction inclusion throughput is collected in the ``Unique ID throughput profiler''. API call latency and success/failure rates are measured by JMeter and the API callback server. VM and container resource utilization data are monitored via AWS CloudWatch. The anchoring performance is measured based on the latency of the transactions writing the Trie of roots content to the tenant chain.

\subsubsection{Results}

\begin{figure}[t]
	\begin{center}
		\center
		\includegraphics[width=0.50\textwidth]{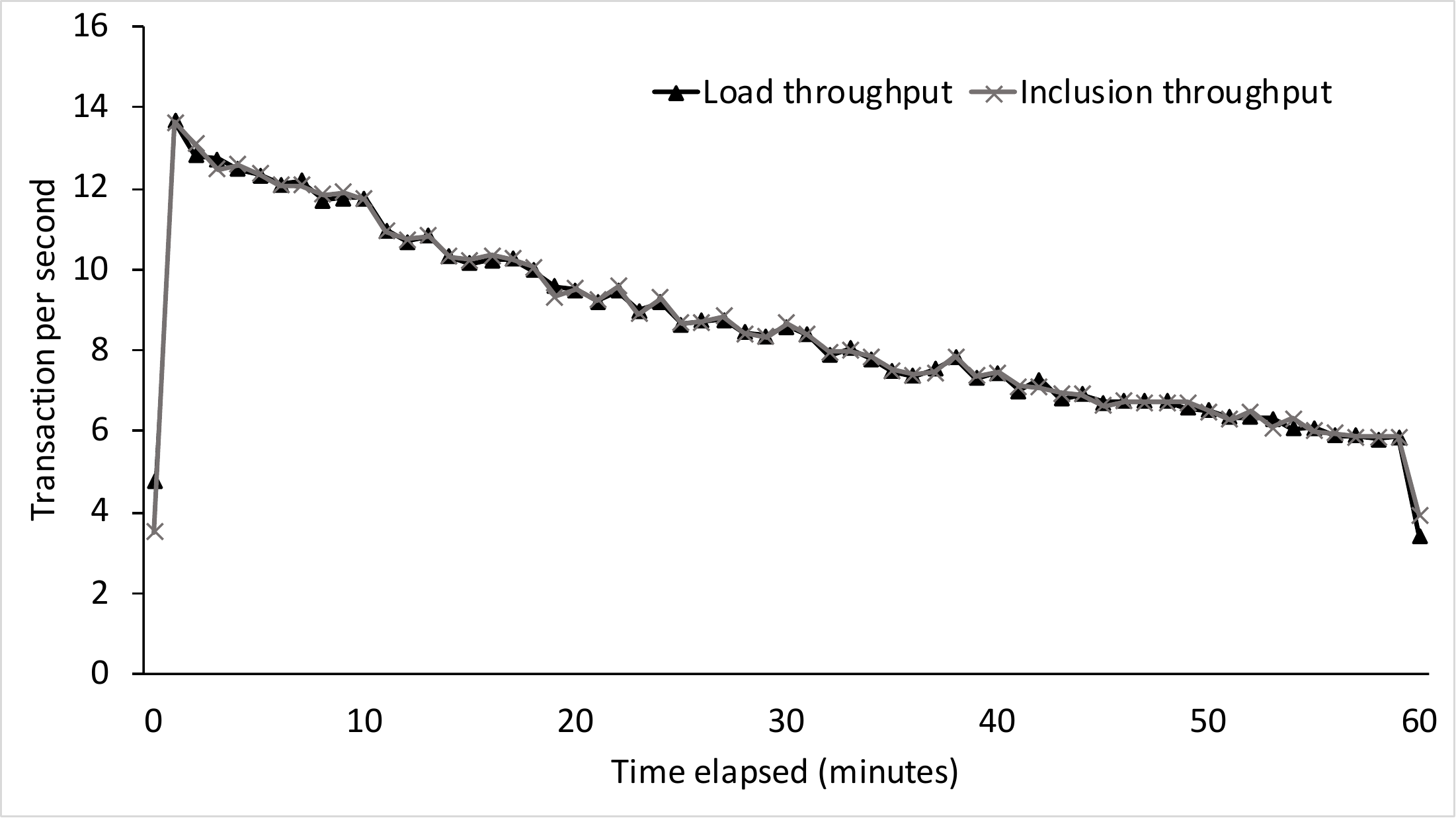}
		\caption{Throughput for Test 1, normal load scenario ($<$15tps).}
		\label{test1}
	\end{center}
\end{figure}

\begin{figure}[t]
	\begin{center}
		\center
		\includegraphics[width=0.50\textwidth]{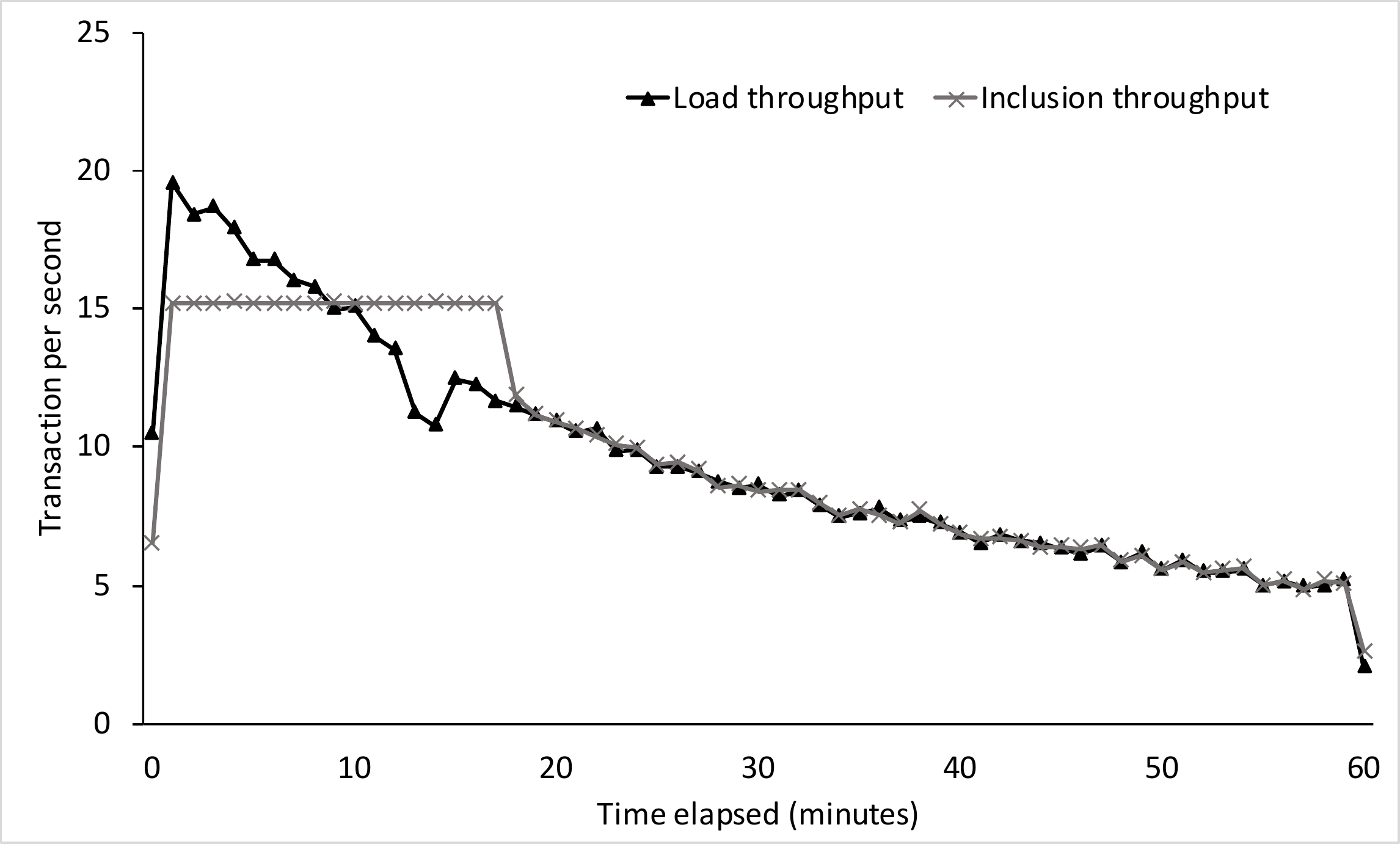}
		\caption{Throughput for Test 2, boundary load scenario (starting at $\approx$18tps).}
		\label{test2}
	\end{center}
\end{figure}

\begin{figure}[t]
	\begin{center}
		\center
		\includegraphics[width=0.50\textwidth]{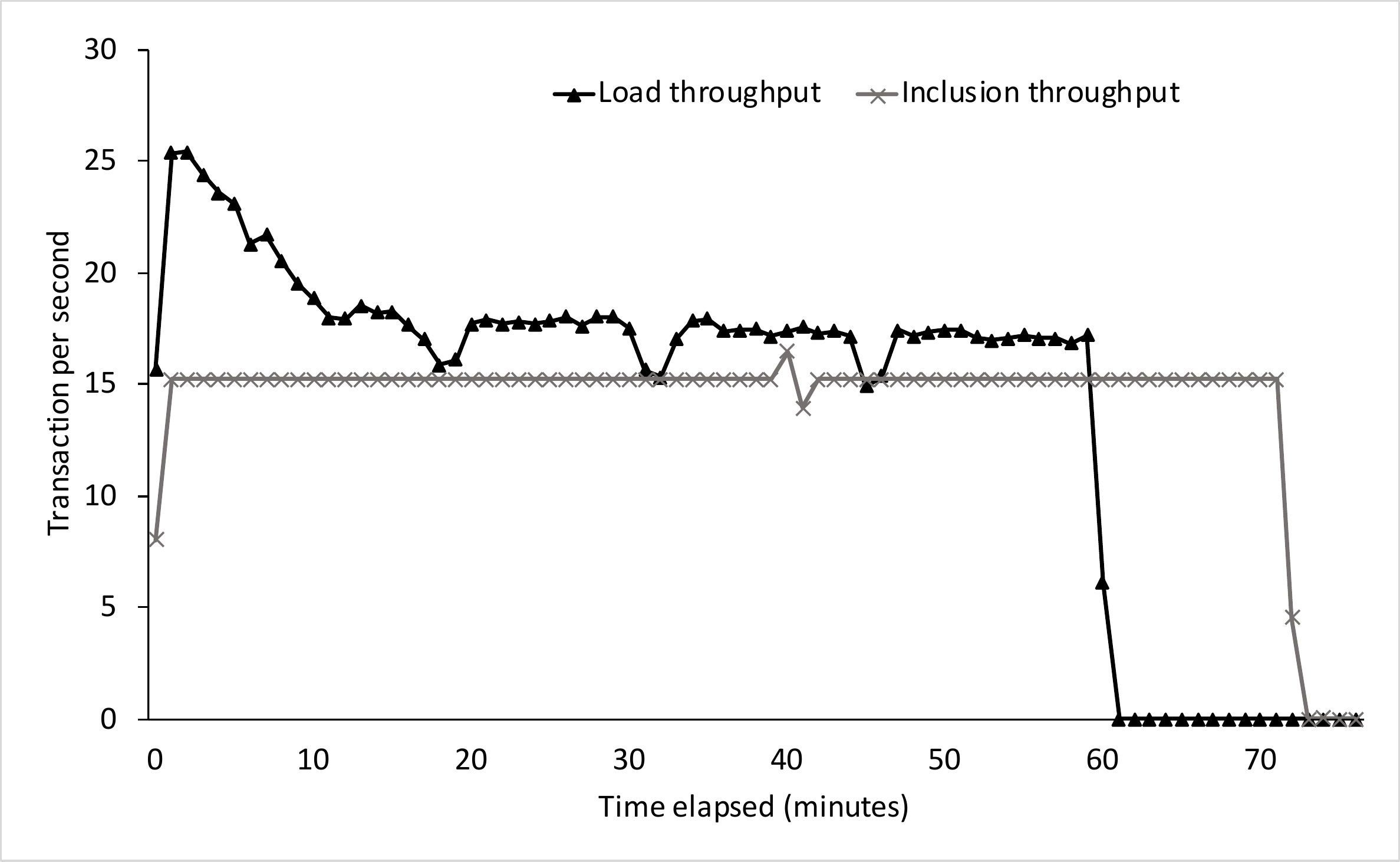}
		\caption{Throughput for Test 3, overload scenario ($\approx$18-25tps).}
		\label{test3}
	\end{center}
\end{figure}

\begin{figure}[t]
	\begin{center}
		\center
		\includegraphics[width=0.50\textwidth]{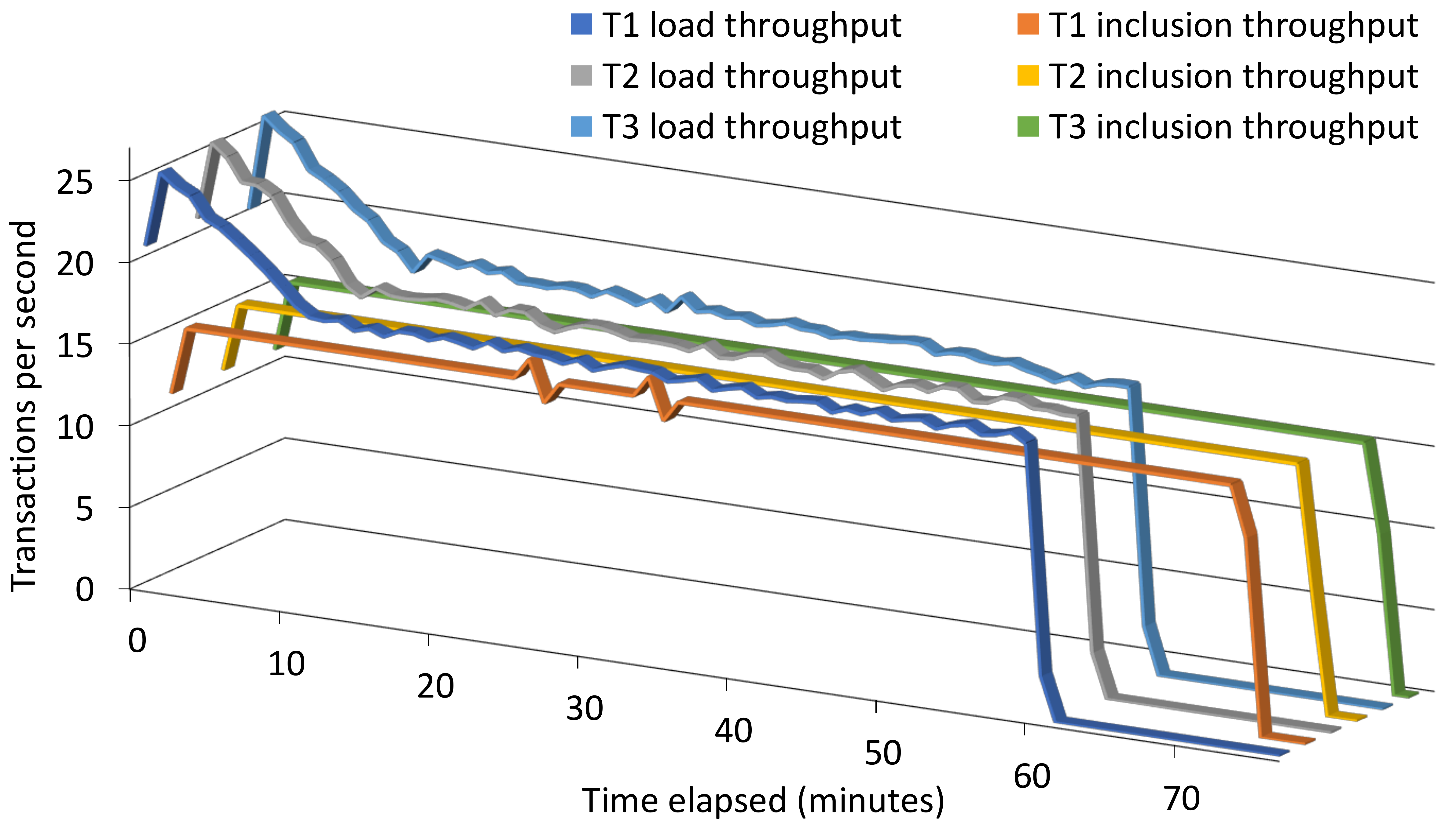}
		\caption{Throughput for Test 4, overload scenario ($\approx$18-25tps).}
		\label{test4}
	\end{center}
\end{figure}

Figures \ref{test1}-\ref{test4} show the throughput measurement results
for Tests 1-4 respectively. The x-axis represents the time
elapsed since the start of the experiment (in minutes), where load is generated from minute 0 to 59.
The y-axis represents the average throughput (tps).

Almost all the transactions are successfully sent and included into the
blockchain without errors. In Tests 1, 2 and 4, no errors occurred. 
In Test 3, 65,506 transactions were sent and 3 errors occurred,
which is reasonable under overload conditions.
Also, in Test 3 and 4, the transactions above the maximum capacity of
the blockchain network were properly queued and eventually included after the generated load finished at the 59-minute mark. Thus, we find that
the implemented prototype can register unique IDs successfully and
efficiently, which meets the first objective of the experiment.

We observed a degradation in performance over the
duration of the first 2 tests, and in the beginning of Test 3 and 4. 
This may be caused by an interplay of the load generator, callback server, and overhead in the trigger implementation, which  
continuously monitors the transaction status after submitting it to Parity.
What can clearly be seen from Test 3 and 4 is that it does not stem from the blockchain, since the inclusion throughput reaches the maximum in minute 1 and (except for a few block-minute-shifts, e.g. around minute 40 in Test 3) stays there until the backlog has been cleared. We also observed from Test 4 that the performance is not impacted by the increased number of tenant chains.

The second objective concerns the performance of anchoring protocol.
Here we measure the total time from start to end of each round (cf.\ \autoref{protocol}).
For all four tests, we measured total times between 9 and 22 seconds per round.
Recall from \autoref{sec:use-case}, that we prioritize anchoring transactions by specifying a higher gas price (fee).
This strategy worked: the anchoring times are not affected
by the load of the tenant chain, even in Test 3 where it is under heavy load. 

Depending on the public blockchain used for anchoring, the total anchoring time can be expected to be dominated by the commit time for the transaction to the public chain. Typical commit times are approx.\ 2-5 minutes for Ethereum, and about 50-100 minutes for Bitcoin\ifHideForReview{\cite{2017-Weber-SRDS}}.

\section{Conclusion and Future Work}
\label{sec:concl}

This paper presents a platform architecture for multi-tenant blockchain systems.
In the design, each tenant is given an individual blockchain, and all tenant chains are anchored to a public blockchain periodically.
The anchoring uses a combined root of all tenant chains, thus achieving data integrity, low cost, and performance and data isolation. 
The proposed architecture has been implemented in a prototype with our industry partner, \laava. 
We evaluate the solution in a three-pronged fashion: by examining requirement fulfilment, by quantitative comparison with two design alternatives, and by quantitative analysis using the prototype. 
The system achieves all objectives.

Although we focused on multi-tenant blockchain-based systems, the proposed  architecture can be applied to many situations requiring multiple blockchains.
Examples include a long-lived and a short-lived blockchain for long and short-running business needs, or a separate blockchain per year.

In future work, we plan to explore the above-mentioned flexible use of anchored chains, as well as the use of other technology platforms -- both for tenant chains and public chains -- with a single anchoring component.


%
\bibliographystyle{abbrv}
\bibliography{ICSA2019}
\end{document}